\documentclass[floatfix,aps,pre,showpacs,showkeys,groupedaddress]{revtex4}
\usepackage{graphicx}
\usepackage{amsmath}
\begin{document}
\title{Distribution of equilibrium free energies in a thermodynamic system with broken ergodicity}
\author{Haijun Zhou and Kang Li}
\affiliation{Institute of Theoretical Physics, the Chinese Academy of Sciences,
  Beijing 100080, China}
\date{July 31, 2007}

\begin{abstract}
  At low temperatures the configurational phase space of a macroscopic
  complex system (e.g., a spin-glass) of $N\sim 10^{23}$ interacting particles
  may split into an exponential number $\Omega_s \sim \exp( {\rm const} \times N)$
  of ergodic
  sub-spaces (thermodynamic states). Previous theoretical studies
  assumed that the equilibrium collective behavior of such a system
  is determined
  by its ground thermodynamic states of the minimal free-energy density,
  and that the equilibrium free energies  follow the distribution of
  exponential decay. Here we show that these assumptions are not necessarily
  valid. For some complex systems, the equilibrium 
  free-energy values may follow a Gaussian distribution
  within an intermediate temperature range, and consequently their equilibrium 
  properties 
  are contributed by {\em excited} thermodynamic states. This work will help 
  improving our 
  understanding of the equilibrium statistical mechanics 
  of spin-glasses and other
  complex systems.
\end{abstract}


\maketitle

\section{Introduction}

A thermodynamic system contains a huge number $N$ of
interacting particles, with $N$ typically 
in the order of $10^{23}$ or larger. The microscopic configurations of such a
system changes with time in a complicated and stochastic manner under the
joint action of internal forces and perturbations from the environment.
At the macroscopic level the 
collective properties of the system are, on the other hand, essentially
time-invariant and can be described
by only a few phenomenological parameters such as the mean energy density and
the  specific heat.
Nevertheless, at certain values of the temperature $T$ or other environmental control
parameters, the macroscopic behavior of the system may also
change abruptly and qualitatively.
Such phase-transition phenomena, being a major research branch of
statistical mechanics for many years, 
are deeply connected with the break down of the ergodicity property of
the system  \cite{Huang-1987,Ruelle-1989}. 

For a large class of complex systems with quenched disorder (heterogeneity) and
frustrations in the interactions among particles as best represented by
spin-glasses \cite{Binder-Young-1986}, when ergodicity breaks down, exponentially
many thermodynamic states will form, each of which corresponds to one ergodic
sub-space of the whole configurational space of the system \cite{Mezard-etal-1987}.
For these systems, it is widely believed (see, e.g.,
Refs.~\cite{Mezard-etal-1987,Rivoire-etal-2004,Castellani-Cavagna-2005,Parisi-2006}) that, 
the {\em equilibrium} properties of the system are determined by the {\em ground} 
thermodynamic states which have the global minimal free-energy density
$f_{\rm min}$, and the distribution of equilibrium
free-energies follows an exponential
law. The {\em excited} thermodynamic states
of free-energy densities $f > f_{\rm min}$
are regarded as irrelevant as long as equilibrium properties are 
concerned, although they dominate the out-of-equilibrium dynamics of
the system (see, e.g., 
\cite{Monasson-1995,Franz-etal-2001,Montanari-RicciTersenghi-2004,Horner-2007}).
For example, a disordered $p$-spin interaction Ising model
($p\geq 3$) \cite{Gross-Mezard-1984,Gardner-1985} is known to have an
ergodic--non-ergodic transition at a temperature $T_{\rm d}$ (the so-called
dynamic transition temperature), but it is expected that the equilibrium
spin-glass phase transition will occur only at a lower temperature $T_{\rm s}$ 
(the static transition temperature). For $T_{\rm d} > T > T_{\rm s}$,
although there are exponentially many thermodynamic states,
all the relevant configurations for the equilibrium properties
are still assumed to reside in the same ergodic
sub-space of the whole configuration space.

In this paper, however, we argue that these 
statements may not necessarily be
correct. Through a general theoretical analysis, we show that the 
equilibrium free-energy densities of an ergodicity-broken system may 
actually follow
a Gaussian distribution with a mean value larger than $f_{\rm min}$. 
Then the equilibrium behavior of the system will be determined by a group of
excited thermodynamic states rather than by the ground thermodynamic states.
Our statement is further supported by analytical and simulation results 
on an exactly solvable model system. 
This work clarifies that, the excited thermodynamic states of a
system of broken ergodicity are important not only to
the dynamical (non-equilibrium) properties of the system
but also to its equilibrium properties.
The theoretical analysis of this paper
may help us to understand more deeply the equilibrium (static) properties
of spin-glasses and
other complex systems. 

When the equilibrium free-energies of an ergodicity-broken
system follows a Gaussian distribution, the ground thermodynamic states of the
system may not be reached by any dynamical process, no matter how long one waits
or which specific cooling schedule is used. In other words, equilibrium studies
based on the Gibbs measure will give a dynamics-independent lower-bound on the
reachably free-energy density. We hope this work will shed light on further
studies of various
fascinating dynamic behaviors of complex systems
\cite{Monasson-1995,Franz-etal-2001,Montanari-RicciTersenghi-2004,Horner-2007,Lunkenheimer-etal-2000}.

\section{General theoretical analysis}

The configuration of a general classical system of
$N$ particles can be
denoted by $\vec{{\bf \sigma}}\equiv\{ \sigma_1, \sigma_2, \ldots, \sigma_N \}$, where the configurational
variable $\sigma_i$ of a particle need not to be discrete or be 
scalar. Each configuration has an energy  ${\cal H}(\vec{\bf \sigma})$.
Starting from an initial configuration, the system evolves with
time and forms a stochastic trajectory in the
whole configurational space $\Gamma$
of the system. At sufficiently high temperatures the system is
ergodic and its trajectory will visit all the (relevant) 
configurations in $\Gamma$ if waited
long enough. More precisely we say a system is ergodic if
two trajectories evolved 
from a pair of randomly chosen initial configurations will, with
probability unity, intersect with each other. In this ergodic situation
the total partition function of the system is expressed as
\begin{equation}
  \label{eq:Z-ergodic}
  Z(\beta)= \sum\limits_{\vec{\bf \sigma} \in \Gamma} \exp\bigl(-\beta {\cal H}(\vec{\bf \sigma}) \bigr) \ ,
\end{equation}
where $\beta\equiv 1/T$ is the inverse temperature. When the system reaches equilibrium, its
free energy is minimized, but its total internal energy still fluctuate
with time. If many measurements are performed on
the internal energy, one will realize that the measured energy values 
follows a Gaussian distribution \cite{Huang-1987,Ruelle-1989}
\begin{equation}
  \label{eq:energy-gaussian}
  \rho(E)= \sqrt{\frac{ \beta^2 }{2 \pi C_E(\beta)}}
  \exp\Biggl( -\frac{ \beta^2 }{2 C_E(\beta)} (E- \langle E \rangle )^2 \Biggr) \ ,
\end{equation}
where $\langle E \rangle$ and $C_E(\beta)$ are, respectively, the
mean total energy and the specific heat of the system. Both $\langle E \rangle$ and
$C_E(\beta)$ are proportional to $N$.

At low temperatures, however, ergodicity may no longer hold.
As the environmental perturbations become weak,
the system may be impossible
to overcome the large free energy barriers between different
regions of the configurational space $\Gamma$; it is then trapped in
one of many ergodic sub-spaces $\Gamma_\alpha$ of $\Gamma$.
In this ergodicity-broken case,  a sub-space $\Gamma_\alpha$
is referred to as a thermodynamic state of the system, which has an
equilibrium free energy $F_\alpha$ as given by
\begin{equation}
  \label{eq:F-alpha}
  F_\alpha(\beta) = - \beta^{-1}  
  \log\Bigl( \sum{_{\vec{\bf \sigma} \in \Gamma_\alpha}} 
  e^{-\beta {\cal H}({\bf \sigma})} \Bigr)\ .
\end{equation}
The energy distribution Eq.~(\ref{eq:energy-gaussian}) still holds in each
thermodynamic state $\alpha$, but now both $\langle E \rangle$ and $C_E(\beta)$ are
thermodynamic state $\alpha$-dependent.

When there are more than one thermodynamic state, the total partition function
Eq.~(\ref{eq:Z-ergodic}) can be re-expressed as a summation over all the thermodynamic
states,
\begin{equation}
  \label{eq:Z-nonergodic}
  Z(\beta)= \sum{_{\alpha}} \exp\bigl(-\beta F_\alpha(\beta)  \bigr) \ ,
\end{equation}
with each thermodynamic state $\alpha$ contributing a term $\exp(-\beta F_\alpha)$.
Equation (\ref{eq:Z-nonergodic}) contains all the information about the
equilibrium properties of an
ergodicity-broken system.
It has the same form as Eq.~(\ref{eq:Z-ergodic}), but with the
configurations $\vec{\bf \sigma}$ being replaced by the thermodynamic 
states $\alpha$. 
This equation indicates that the contribution of a
thermodynamic state $\alpha$ to the equilibrium property of the system is
proportional to $\exp(-\beta F_\alpha(\beta))$. Although such a 
Gibbs measure is arguably not holding in an out-of-equilibrium 
dynamics, it is commonly used in equilibrium studies. In this work we
also stick to this Gibbs measure. 

To further understand this Gibbs measure, in this paragraph we try to
give an interpretation based on a gedanken dynamical process of
heating and annealing (but we emphasize that the results of this paper is
independent of this interpretation). 
For the
system to escape a thermodynamic state $\alpha$, a large external perturbation
has to be applied. This might be achieved by first heating the system
and then cooling it \cite{Zhou-2007a,Zhou-2007b}.
As the system is heated to a high temperature, it becomes ergodic and memory
about its prior history is lost. After the system is cooled down
slowly to its original low temperature, it may reach a different thermodynamic
state $\alpha^\prime$ at the end of this process. (During the annealing 
process of this gedanken experiment, the system
may be driven by a global and parallel dynamical rule.) 
All the thermodynamic states of the system at a low temperature $T$ 
will therefore be explored if one repeats extremely many times
this heating-annealing experiment. With this external assistance, the system
again becomes ergodic at the level of thermodynamic states.
Since the prior history of the system is completely destroyed in
the heating-annealing experiment, the frequency of the system
reaching a thermodynamic state
$\alpha$ supposed to be
given by the Gibbs measure $e^{-\beta F_\alpha}/ Z(\beta)$. 

Let us denote by $\Omega_{{\rm s}}(F)$ the total number of thermodynamic states in the
system with free energy $F$. Then the equilibrium free energy distribution
is governed by 
\begin{equation}
  \label{eq:F-profile}
  P(F) \propto \Omega_{\rm s}(F) e^{-\beta F} = 
  \exp\bigl( - \beta  F + S_{\rm s}(F) \bigr) \ ,
\end{equation}
where, $S_{\rm s}(F)= \log \Omega_{\rm s}(F)$ is the entropy at the level of 
thermodynamic states. $S_{\rm s}(F)$ is a concave and increasing function of $F$.
We are interested in systems with exponentially many thermodynamic states,
i.e., systems with $S_{\rm s}(F)$ being proportional to the size $N$ in leading
order.

If at the minimal free energy $F_{\rm min}(\beta)$, the first derivative of 
$S_{\rm s}(F)$ 
is greater than $\beta$, i.e., $S^\prime_{\rm s}(F_{\rm min}) > \beta$, 
there exists a free energy
value $F=\overline{F} > F_{\min}(\beta)$ such that
$S^\prime_{\rm s}(\overline{F}) = \beta$. 
At the vicinity of $\overline{F}$, the entropy $S_{\rm s}(F)$ is expressed as
\begin{equation}
  \label{eq:S-s-expand}
  S_{\rm s}(F)= S_{\rm s}(\overline{F}) + \beta (F- \overline{F} )
  - \frac{ \beta^2 }{2 C_{F}(\beta)} (F- \overline{F} )^2 \ .
\end{equation}
After inserting Eq.~(\ref{eq:S-s-expand}) into Eq.~(\ref{eq:F-profile}) we find that,
at equilibrium, the probability of being in a state of free energy $F$ is governed 
by the following Gaussian distribution
\begin{equation}
  \label{eq:r-F-1}
  P(F) =  \sqrt{\frac{ \beta^2 }{2 \pi C_F(\beta)}}
  \exp\Biggl( -\frac{ \beta^2 }{2 C_F(\beta)} (F- \overline{F})^2 \Biggr) \ .
\end{equation}
From Eq.~(\ref{eq:r-F-1}) it is clear that
$\overline{F}$ is the mean free energy value of the equilibrium thermodynamic
states, and $C_F(\beta)\propto N$ characterizes the fluctuation of the
equilibrium free energies. 
Since $\overline{F}(\beta) > F_{\rm min}(\beta)$, we conclude that the
equilibrium properties of the system at inverse temperature $\beta$ are
determined by those excited thermodynamic states whose free energy density
$f(\beta)=\overline{F}/N$ is larger than the minimal free energy density
$f_{\rm min}(\beta)=F_{\rm min}/N$. The ground thermodynamic states of
free energy density $f_{\rm min}(\beta)$ actually do not
contribute to the equilibrium properties of the system.

On the other hand, if the entropy $S_{\rm s}(F)$ has the property that at 
$F=F_{\rm min}(\beta)$ its first derivative is less than $\beta$, i.e.,
\begin{equation}
  S^\prime (F_{\rm min}) = x \beta
\end{equation}
with $0 \leq x < 1$,
then Eq.~(\ref{eq:F-profile}) suggests that 
the equilibrium free energies will follow an exponential low:
\begin{equation}
  \label{eq:r-F-2}
  P(F) \propto  e^{- \beta(1-x) (F-F_{\rm min}(\beta) )} \ , \;\;\; F \geq F_{\rm min}(\beta) \ . 
\end{equation}
Consequently, the equilibrium properties of the system will be
contributed by the ground thermodynamic states of free
energy density $f_{\rm min}(\beta)$; and the fluctuation of the observed free
energies is only of order unity.

\section{Grand Free Energy}

To treat the two free-energy distributions 
of the preceding section 
with the same mathematical framework, we need to define a grand
free energy for the system. Following the work of 
M{\'{e}}zard, Parisi, and Zecchina 
\cite{Mezard-etal-2002,Mezard-Parisi-2003} on the mean-field theory of
$T=0$ spin-glasses, we can decouple microscopic configurations and
macroscopic states by  introducing an artificial
inverse temperature $y$ at  the level of thermodynamic states.
The system's grand free energy $G(\beta; y)$ \cite{Zhou-2007b} is defined by
\begin{eqnarray}
  G(\beta; y) &\equiv& - y^{-1}
  \log\Bigl( \sum{_{\alpha}} e^{-y F_\alpha( \beta ) } \Bigr) 
  \label{eq:grand-free-energy} \\
  &= &- y^{-1} \log\Bigl[ \int {\rm d} f e^{N 
    \bigl( \Sigma(f)- y f \bigr)}   \Bigr]  \ .
  \label{eq:grand-free-energy-2}
\end{eqnarray}
In the thermodynamic limit of $N \to \infty$, the grand
free energy density is 
\begin{equation}
  \label{eq:gfe-density}
  g(\beta;y)\equiv \lim\limits_{N\to\infty} \frac{G(\beta;y)}{N} \ .
\end{equation}
In Eq.~(\ref{eq:grand-free-energy-2}), $\Sigma(f)\equiv S_{\rm s}(N f)/ N$ measures the
entropy density at the level of thermodynamic states; it is called 
the complexity of the system at free energy density $f$
\cite{Mezard-Parisi-2003}.
The adjustable
parameter $y$ controls which thermodynamic states will contribute to the
grand free energy $G(\beta;y)$.
Equation~(\ref{eq:grand-free-energy-2}) indicates that, 
when the re-weighting parameter $y$ is
not too large, the grand free energy is contributed by the excited
thermodynamic states of
free energy density satisfying $\Sigma^\prime(f)= y$. The relevant free energy
 density and complexity are 
related to the grand free energy density by
\begin{eqnarray}
  f(\beta; y) &=& \frac{ \partial y g(\beta; y) }{\partial y} \ ,   \label{eq:fe} \\
  \Sigma(\beta;y) &=& y^2 \frac{\partial g(\beta;y)}{\partial y} > 0\ .   \label{eq:cpl}
\end{eqnarray}
On the other hand, 
when $y > y^*(\beta) \equiv \Sigma^\prime\big(f_{\rm min}(\beta) \bigr)$, the grand free energy
is contributed by the ground thermodynamic states of the system, therefore
\begin{eqnarray}
  f\bigl(\beta; y > y^*(\beta) \bigr) &=& f_{\rm min}(\beta) \label{eq:fe2} \\
  \Sigma\bigl(\beta; y > y^*(\beta) \bigr) &=& 0 \label{eq:cpl2}
\end{eqnarray}

From Eq.~(\ref{eq:cpl}) and (\ref{eq:cpl2}) we know that, (1) the minimal free energy
density $f_{\rm min}(\beta)$ corresponds to $y=y^*(\beta)$, where the complexity
$\Sigma(\beta;y)$ drops to zero; 
(2) if $\Sigma(\beta;\beta) > 0$, then
$f(\beta; \beta)> f_{\rm min}(\beta)$ is the mean free energy density of the
thermodynamic states which dominate the equilibrium properties of the system.

\section{Results on the $p$-spin interaction Ising spin-glass model}

Let us complement the above-described general
analysis with a concrete example, namely the 
$p$-spin interaction Ising model on a complete
graph \cite{Gardner-1985}. The Hamiltonian of the model is
\begin{equation}
  \label{eq:P-spin-model}
  {\cal H}({\bf \sigma}) = 
  - \sum\limits_{1\leq i_1 < \ldots < i_p \leq N} J_{i_1 i_2 \ldots i_p} \sigma_{i_1}
  \sigma_{i_2} \ldots \sigma_{i_p} \ ,
\end{equation}
where the spin variables $\sigma_{i}=\pm 1$ 
and the quenched (time-independent) coupling
constant $J_{i_1 \ldots i_p}$ is identically and independently distributed 
according to
\begin{equation}
  \label{eq:J-form}
  \omega(J_{i_1 i_2 \ldots i_p} ) = \sqrt{\frac{N^{p-1}}{ \pi p! J^2 }} \exp\Biggl(
  -\frac{N^{p-1}}{p! J^2} J_{i_1 i_2 \ldots i_p}^2
  \Biggr)
\end{equation}
with $J$ being a constant parameter (the energy unity of the system).
For $p=2$, Eq.~(\ref{eq:P-spin-model}) is
the celebrated Sherrington-Kirkpatrick model
\cite{Sherrington-Kirkpatrick-1975,Mezard-etal-1987}. For $p\geq 3$, earlier efforts
\cite{Gardner-1985,Rivoire-etal-2004} have found that
the system has two transitions, a dynamic transition followed
by a lower-temperature static transition. The dynamic transition is related to
the onset of ergodicity-breaking and is important for out-of-equilibrium
processes, but it was not regarded as a real equilibrium phase-transition.

\begin{figure}[ht]
  \includegraphics[width=0.85\linewidth]{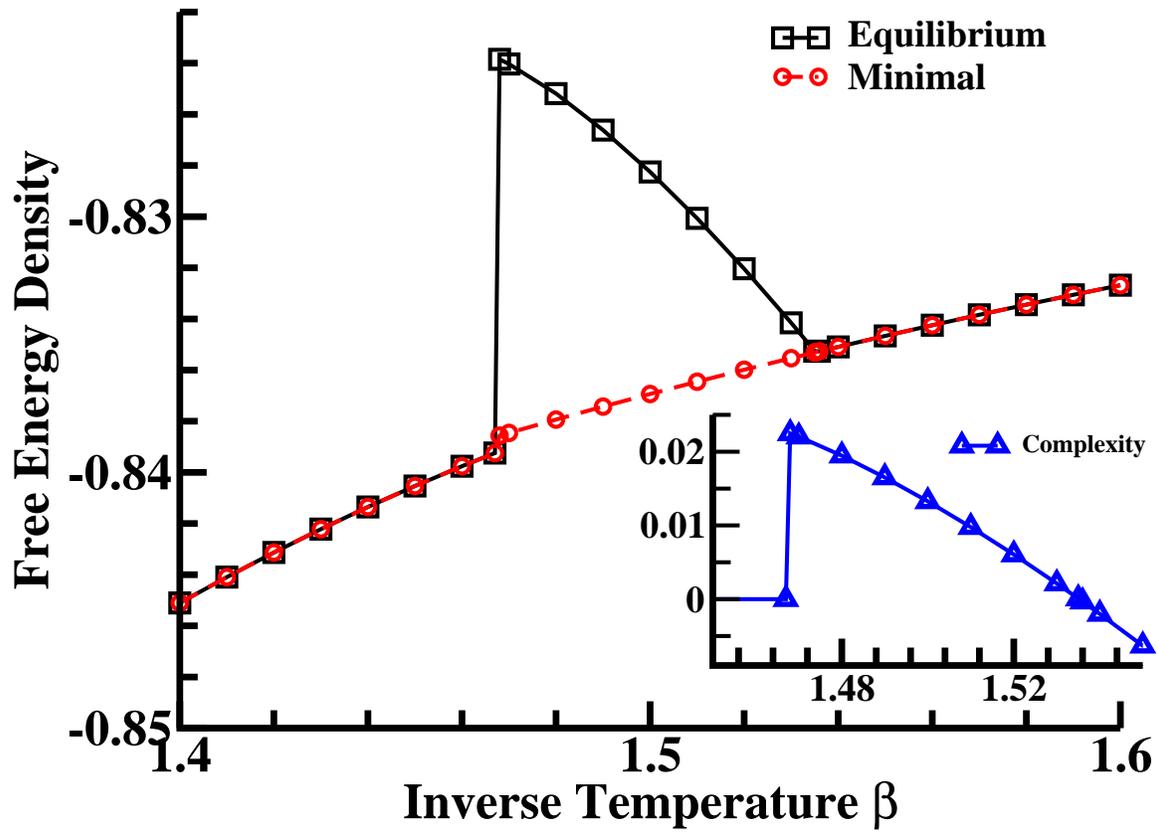}
  \caption{\label{fig:free-energy}
    The mean equilibrium free energy density and the 
    minimal free energy density of the $3$-spin interaction
    Ising model Eq.~(\ref{eq:P-spin-model}) on a complete graph of $N=\infty$. 
    Inset shows the
    complexity $\Sigma(\beta;\beta)$ as a function of $\beta$. 
    For $\beta \in (1.468,1.5352)$ the
    equilibrium properties of the system are determined by excited thermodynamic
    states.}
\end{figure}

If we assume that all the thermodynamic states of the model system
Eq.~(\ref{eq:P-spin-model}) are evenly distributed in the whole configurational space
$\Gamma$, i.e., there is no further clustering of the thermodynamic states, the grand
free-energy density of the system as defined by Eq.~(\ref{eq:gfe-density})
can be obtained through the cavity method \cite{Mezard-etal-1987} (see
also \cite{Zhou-2007b}). The final expression for $g(\beta; y)$ is
%
\begin{eqnarray}
 & &  g(\beta; y)  =  -\frac{1}{\beta} \log 2 
   -  \frac{p-1}{4} J^2 (y q_0^p + (\beta-y) q_1^p) \nonumber \\
  & & \;\;\; - \frac{1}{4} \beta J^2 (1- p q_1^{p-1})  
 - \frac{1}{y} \int \frac{ {\rm d} z_0}{ \sqrt{\pi}} e^{-z_0^2} 
  \log\Bigl[ \int \frac{ {\rm d} z_1}{\sqrt{ \pi}} e^{-z_1^2} \nonumber \\
  & & \;\;\; \times \cosh^{y/\beta} (
  \beta J \lambda_0 z_0 + \beta J \lambda_1  z_1 ) \Bigr]
  \ ,
  \label{eq:gfe-density2}
\end{eqnarray}
where $\lambda_0=\sqrt{p} q_0^{(p-1)/2}$,
$\lambda_1= \sqrt{p} ( q_1^{p-1}-q_0^{p-1})^{1/2}$,
$q_0 = \langle m \rangle^2$,  and $q_1= \langle m^2 \rangle$, with $m$ 
being the magnetization of a
vertex in one thermodynamic state, and $\langle \cdots \rangle$ means 
averaging over all
the thermodynamics states $\alpha$ of the system (each of them is
weighted with
the factor $e^{-y F_\alpha(\beta)}$). 
$q_0$ and $q_1$ satisfy $\partial g/ \partial q_0 = \partial g / \partial q_1 = 0$.
Equation (\ref{eq:gfe-density2}) was first derived in \cite{Gross-Mezard-1984}
using the replica trick, and was regarded as the free-energy density of
the system \cite{Gross-Mezard-1984,Gardner-1985}. But we see that actually $g(\beta;y)$ is
the grand free-energy density, which combines
both the free energy effect and the entropy effect (at the level of thermodynamic
states) of the system. 

\begin{figure}[ht]
  \includegraphics[width=0.85\linewidth]{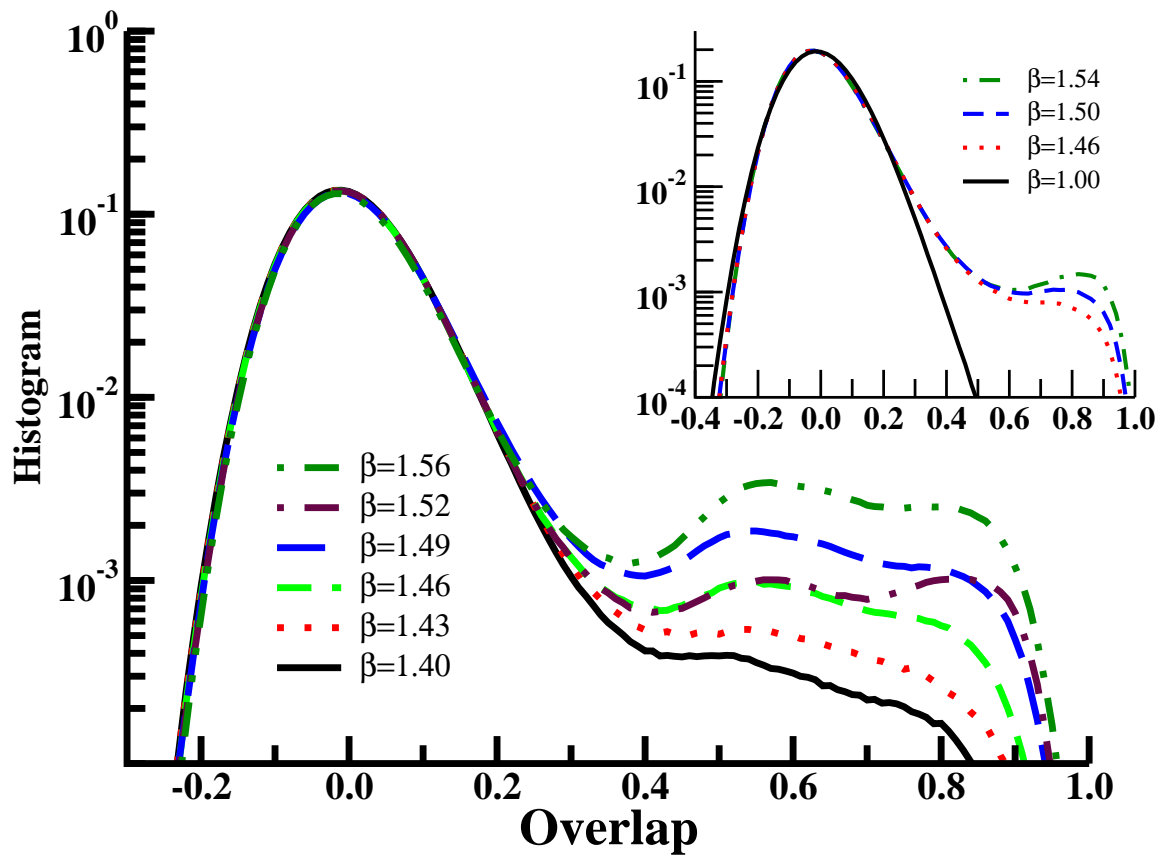}
  \caption{
    \label{fig:overlap}
    Overlap histograms for a
    $3$-spin interaction Ising systems of $N=200$ vertices  (the main figure)
    and $N=100$ vertices (the inset). 
    Different curves correspond to different inverse temperatures.  
  }
\end{figure}

For an infinite system with $p=3$, the mean values of the equilibrium and the minimal 
free energy density are shown in Fig.~\ref{fig:free-energy} 
as a function of the inverse temperature $\beta$. Ergodicity of 
the system breaks down
at $\beta_{1}\simeq 1.468$, where the whole configuration space splits into 
exponentially
many ergodic sub-spaces. The equilibrium and the minimal free energy density of 
the system has a
jump at $\beta_1$, but the energy and grand free-energy densities
are both continuous at this point. For $\beta_1 < \beta < \beta_2\simeq 1.5352$,
the mean equilibrium free-energy density is higher 
than the minimal free-energy density (which is obtained by setting
$y > \beta$), and the complexity of the system decreases continuously with
$\beta$ and drops to zero at
$\beta_2$. For $\beta > \beta_2$, the mean equilibrium
free-energy density is identical to the minimal
free energy density of the system. The above-mentioned results also hold when
one considers the possibility of further clustering of the thermodynamic states
or splitting of each thermodynamic state into sub-states 
\cite{Gardner-1985,Montanari-RicciTersenghi-2003}.

For a system with small size $N$ ergodicity will be preserved even at low
temperatures; but the relevant configurations of the system may show some degree of
clustering. To detect this organization, we can
calculate the overlaps between the sampled independent configurations of the
system. The overlap of two configurations $\vec{\bf \sigma}^1$ and
$\vec{\bf \sigma}^2$ is defined as  \cite{Mezard-etal-1987}
\begin{equation}
  \Lambda_{12}= \frac{1}{N} \sum\limits_{j=1}^{N} \sigma_j^1 \sigma_j^2 \ .
\end{equation}
The overlap histograms for two finite systems of sizes $N=100$ and
$N=200$ 
are shown in Fig.~\ref{fig:overlap}. Two peaks appear in the histograms when
$\beta$ approaches the theoretically predicted value $\beta_1$. 
The peak at $\Lambda \simeq 0$ 
is due to pairs of configurations from different domains of the 
configurational
space, and the other peak at $\Lambda \simeq 0.8$ (for $N=100$) or
$\Lambda \simeq 0.6-0.8$ (for $N=200$)
corresponds to the overlaps between 
configurations from the same domain of the configurational space.
Figure~\ref{fig:overlap} also demonstrates that, as the system size $N$ increases,
the organization of the configurational phase space becomes more complex.

\section{Conclusion and discussion}

In this paper we studied the equilibrium
properties of a thermodynamic system with broken ergodicity such as a spin-glass.
If the number of thermodynamic states increases exponentially fast with
the system size $N$ at low temperatures, we show that the equilibrium
free-energy distribution of the system may be Gaussian, and consequently the
equilibrium static properties of the system are determined by excited 
thermodynamic states
of the system, whose free-energy densities are higher than the minimal free-energy
density of the system. A grand free energy function (with an adjustable
parameter $y$) was defined in this paper following
the earlier work of Refs.~\cite{Mezard-Parisi-2003,Mezard-etal-2002} 
to calculate the mean value of the
equilibrium free-energy density and the complexity of the system.

The mean-field theory of spin-glasses by Parisi and colleagues
\cite{Mezard-etal-1987,Parisi-2006} was based on the assumption that the
equilibrium free-energies of the system obey an exponential distribution.
Under that theory, only the thermodynamic states of the ground
free-energy density are {\em allowed}
to contribute to the equilibrium properties of the
system. As we now know, for disordered systems with two-body interactions
\cite{Sherrington-Kirkpatrick-1975} this assumption of exponential-distribution is
valid. But for a system with many-body interactions, there may exist a temperature
window within which the free-energy distribution is Gaussian. In this later case,
Fig.~\ref{fig:free-energy} demonstrates that the mean value of
the equilibrium free-energy densities
{\em decreases} with temperature.  This apparently will cause an
entropy crisis, but actually the entropy of a thermodynamic state is positive.
Notice that when the free-energy distribution is Gaussian, different groups of
thermodynamic states are taking the dominant role as the temperature changes.
The predictions of the present work can be further 
checked by Monte Carlo simulations
on a large finite-connectivity complex system with many-body interactions. 

In this work, we focused on the equilibrium statical properties of an
ergodicity-broken system and assumed that the significance of
each thermodynamic state $\alpha$ is proportional to $\exp(-\beta F_\alpha)$, with
$F_\alpha$ being its free energy. This assumption may not be valid for
out-of-equilibrium dynamical processes. For these later non-equilibrium processes,
it has been suggested that
the system will typically be 
trapped to a free energy level which corresponds to the maximal
complexity. When the system is cooled down slowly from a high temperature,
the reachable thermodynamic states depend strongly 
on the specific dynamical rules
used \cite{Montanari-RicciTersenghi-2004,Horner-2007}. The mean equilibrium
free energy density discussed in this paper, although may not being achievable
in a dynamical experiment, sets a lower-bound on the dynamically reachable 
free energy density. As demonstrated by Fig.~\ref{fig:free-energy}, in an intermediate
temperature range, 
this lower bound may be well above the minimal free energy density of the system.

\acknowledgments

{H.Z.} acknowledges the hospitality of 
Pik-Yin Lai and other colleagues at the Physics Department of the
National Central University, where this work was finished. The simulation 
of Fig.~\ref{fig:overlap} was performed in the PC clusters of the State Key
Laboratory for Scientific and Engineering Computing (Beijing).

%
%


\end{document}